\documentstyle[12pt,aasms4]{article}

\def\be{\begin{equation}}
\def\ee{\end{equation}}
\def\ba{\begin{eqnarray}}
\def\ea{\end{eqnarray}}

\def\Atil{\mbox{\~{A}}}
\def\Ncal{{\cal N}}

\def\la{\mathrel{\mathpalette\fun <}}

\def\fun#1#2{\lower3.6pt\vbox{\baselineskip0pt\lineskip.9pt
        \ialign{$\mathsurround=0pt#1\hfill##\hfil$\crcr#2\crcr\sim\crcr}}}

\slugcomment{To be submitted to ApJ}

\begin{document}
\null\vspace{-62pt}
\begin{flushright}
September 29, 1997\\
To appear in ApJ on May 1, 1998
\end{flushright}

\title{Implications of Cosmic Microwave Background \\
Anisotropies for Large Scale Variations in Hubble's Constant}

\author{Yun Wang, David N. Spergel, and Edwin L. Turner}
\affil{{\it Princeton University Observatory} \\
{\it Peyton Hall, Princeton, NJ 08544\\}
{\it email: ywang,dns,elt@astro.princeton.edu}}

\vspace{.4in}
\centerline{\bf Abstract}
\begin{quotation}

Low amplitude (linear regime) cosmic density fluctuations lead
to spatial variations in the locally measurable value of $H_0$
(denoted as $H_L$), $\delta_H \equiv (H_L-H_0)/H_0$,
which are of order $3-6$\% (95\% confidence interval)
in a sphere of 200$\,h^{-1}$Mpc in diameter,
and of order $1-2$\% in a sphere of 400$\,h^{-1}$Mpc in diameter,
for three currently viable structure formation models
(tilted CDM, $\Lambda$CDM, and open CDM)
as normalized by the 4 year COBE DMR data.

However, the true matter distribution power spectrum may differ from
any of the currently viable models. For example, it may
contain sharp features which have escaped detection so far.
The measured CMB dipole velocity (the Galaxy's peculiar velocity 
with respect to the CMB rest frame) provides additional constraints
on the probability distribution of $\delta_H$ that supplement our limited
knowledge of the power spectrum.
For a matter power spectrum which consists of
the smooth power spectrum of a viable cosmological model
plus a delta-function bump, we find that given the CMB dipole velocity,
the 95\% CL upper limit of $|\delta_H|$ increases approximately by a 
factor of two, but the probability distribution of $\delta_H$ 
is non-Gaussian, with 
increased probability at small $\delta_H$ compared to Gaussian.
Abandoning model power spectra entirely, we find that the observed
CMB dipole velocity alone provides a very robust limit,
$\sqrt{ \langle \delta_H^2 \rangle_R } <10.5\, h^{-1}\mbox{Mpc}/R$
at 95\% CL, in a sphere of radius $R$, for an arbitrary power
spectrum.

Thus, variations between currently available local measures of $H_0$ 
and its true global value of a few to several percent are to be expected 
and differences as large as 10\% are possible based on our current 
knowledge of the CMB anisotropies.

\end{quotation}


\section{Introduction}

In the standard gravitational instability scenario (\cite{Peebles80,Peebles93})  for cosmic structure
formation, linear growth of density fluctuations is produced by, and
produces, spatial variations of the expansion rate.  In this rather
generic scenario, such variations are the inevitable implication of the
extremely large scale structures which have been detected in the galaxy
distribution, primarily via redshift surveys (e.g., \cite{Dacosta94,Lauer94,Lin96,Tadros96}).

The connection of such expansion rate variations, which are often thought
of as peculiar velocity fields, to the large scale density distribution is
of direct interest (\cite{Strauss95}) but is also potentially
important as a source of systematic error in attempts to measure Hubble's
constant $H_0$, the overall mean expansion rate of the Universe.  
In particular, if the expansion rate is correctly measured in a local
volume which is not sufficiently large compared to the biggest significant
cosmic structures, the strong possibility of a difference between it and the
true cosmic value of $H_0$ must be considered.

This potential problem has long been known in principle, was emphasized
in the context of modern structure formation models by 
Turner, Ostriker, \& Cen (1992)
and has subsequently been considered by several authors
(\cite{Wu95,Nakamura95,Nakao95,SWD,Wu96,Shi97}).  The issue is of greater
interest than ever at present for two reasons:  First, measurements of
$H_0$ by conventional distance ladder techniques have improved dramatically
and now achieve credible precisions of order 10 to 20 percent.  It follows
that a systematic difference between local and global expansion rates of a
comparable fractional size are important sources of error.  Second, there is
suggestive, though not yet compelling, evidence that direct physical methods
for measuring $H_0$ on extremely large scales (out to redshifts of order
unity) (\cite{Birkinshaw91,Jones93,Schechter97,Kundic97}) give a somewhat smaller value than
the best distance ladder determinations which apply out to redshifts of a
tenth or less (\cite{Graham97,Tonry97,Eastman96}).

In this paper we attempt to exploit our best and most direct empirical
information about large scale (linear) cosmic density fluctuations, namely
observed anisotropies in the cosmic microwave background (CMB), in order
to predict and/or constrain the expected resulting variations in the
expansion rate.  Specifically, we study these intrinsic fluctuations, 
$\delta_H \equiv (H_L-H_0)/H_0$, where $H_L$ is the local measure of $H_0$,
in terms of the matter distribution power spectrum $P(k)$.
(Note that although galaxies are biased tracers of the mass density field,
on the large scales which are of interest to this paper $-$ 100$\,$Mpc
or larger, linear biasing is a reasonable assumption for
conventional structure formation scenarios.)

In fact, we present here three separate calculations of this sort, proceeding
from the one which gives the strongest result but which is most model
dependent to the one which is weakest but most robust (model independent).
First, we simply base our input $P(k)$ on standard structure formation models
in terms of its shape and
use CMB observations only to set the overall normalization or amplitude.
Here, our results agree with recent work by Shi \& Turner (1997).
Second, we consider an input $P(k)$ with a shape and amplitude constrained by
CMB fluctuations; for some scales ($k$ values) $P(k)$ is effectively directly
observed, but we allow for the possibility of strong features on those scales
not directly sampled by currently available data by imposing limits based on
the CMB dipole (the Galaxy's peculiar velocity with respect to the
CMB rest frame).  Third, we impose no 
constraints at all on the input $P(k)$ (which does not even appear explicitly
in this calculation) other than that it be Gaussian and not violate the 
aforesaid CMB dipole constraint.  

Section 2 contains general expressions for $\delta_H$ and
related variables.
In Section 3, we compute $\delta_H$ for matter power spectra given
by three viable cosmological models which are normalized by 
the 4 year COBE DMR data and satisfy constraints from large scale 
structure data.
In Section 4, we study the effects of unexpected features in $P(k)$ which
may boost $\delta_H$ by adding a delta-function bump
to a smooth matter power spectrum given by a viable cosmological model.
In Section 5, we apply Bayesian statistics to derive robust upper
limits on $\delta_H$ and related variables, using the CMB dipole velocity 
of $v$=627 km/s (\cite{Kogut93,Fixsen94}); these upper limits are 
independent of the actual form of the matter power spectrum.
Section 6 contains discussions and a summary.

\section{General Expressions}

For an observer at $\bf{r}_i$ who measures the Hubble's constant by
summing over the radial recession velocity divided by distance of
$N$ objects located at $\bf{r}_j$, ($j=1$, 2, ..., $N$) (\cite{Turner92})
\be
\delta_H({\bf r}_i) \equiv \frac{H({\bf r}_i)-H_0}{H_0}
= \frac{1}{N} \sum_{j\neq i}
\frac{ {\bf v}_j  \cdot ({\bf r}_j-{\bf r}_i) } {H_0 |{\bf r}_j-{\bf r}_i|^2}.
\ee
To find $\delta_H$ for a sphere of radius $R$ centered around ${\bf x}$,
we write (\cite{SWD})
\be
\delta_H({\bf x})^R = \int d^3{\bf y} \,\frac{{\bf v}({\bf y})}
{H_0} \cdot \frac{ {\bf y}-{\bf x} }{|{\bf y}-{\bf x}|^2} \, 
W({\bf y}-{\bf x}),
\ee
where $W({\bf y}-{\bf x})$ is the top hat window function with radius $R$,
\be
W({\bf y}-{\bf x})= \left\{ \begin{array}{ll} (4\pi R^3/3)^{-1}, &
\hskip 1cm |{\bf y}-{\bf x}| \leq R \\
0, & \hskip 1cm |{\bf y}-{\bf x}| > R \end{array} \right.
\ee
In linear perturbation theory (Peebles 1993, 1980), the Fourier
component of ${\bf v}({\bf y})=(2\pi)^{-3} \int d^3 {\bf k}\,
{\bf v}_k\, \exp(-i {\bf k}\cdot {\bf y})$ is
\be
{\bf v}_k = \frac{\Omega_0^{0.60} H_0}{i k}\, \delta_k\, \hat{k},
\ee
for an open or flat universe.
$\delta_k$ is the Fourier component of the density fluctuation.
Hence
\be
\delta_H({\bf x})^R = \frac{\Omega_0^{0.60}}{(2\pi)^3}
\int d^3{\bf k}\, \delta_k\, {\cal L}(kR) \, e^{-i{\bf k}\cdot {\bf x}},
\ee
where
\be
{\cal L}(x) = \frac{3}{x^3} \left( \sin x - \int_0^x dy \, 
\frac{\sin y}{y} \right).
\ee
Note that ${\cal L}(x \rightarrow 0) =-1/3$, and ${\cal L}(x)<0$,
as expected; overdensities ($\delta_k>0$) lead to infall, which
leads to the underestimate of the Hubble's constant ($\delta_H<0$).

$\delta_H$ is a Gaussian random variable, with mean 0 and
variance
\be
\label{eq:<dH^2>}
\langle \delta_H^2\rangle_R = \frac{\Omega_0^{1.20}}{2\pi^2 R^2}
\int_0^{\infty}dk\, P(k) \left[(kR) {\cal L}(kR)\right]^2.
\ee
The power spectrum $P(k)=|\delta_k|^2$. 
The variance of density fluctuations in a sphere of radius R is
\be
\label{eq:<drho^2>}
\left\langle \left(\frac{\delta\rho}{\rho}\right)^2 
\right\rangle_R = \frac{1}{2\pi^2 R^2}
\int_0^{\infty}dk\, P(k) \left\{(kR) \, \left[\frac{3j_1(kR)}{kR}
\right]\right\}^2.
\ee
Clearly, the fluctuations in the measured expansion rate stem directly 
from the fluctuations in matter density. 

We can define $\Omega_L \equiv 8\pi G \rho_L/(3 H_L^2)$. To lowest order
in $\delta\rho/\rho$ and $\delta_H$, $\delta_{\Omega} \equiv
(\Omega_L-\Omega_0)/\Omega_0 = \delta\rho/\rho - 2\delta_H$. We find
\be
\label{eq:<dOmega^2>}
\langle \delta_{\Omega}^2 \rangle_R = \frac{1}{2\pi^2 R^2}
\int_0^{\infty}dk\, P(k) \left\{(kR) \left[\frac{3j_1(kR)}{kR}
-2\Omega_0^{0.60}{\cal L}(kR)\right] \right\}^2.
\ee
Note that the fluctuations in $\Omega_L$ are contributed additively
by $\delta\rho/\rho$ and $\delta_H$ (${\cal L}(x)<0$).
$\Omega_L=8\pi G \rho_L/(3 H_L^2)$ is related but not identical 
to the locally measured $\Omega$ (by counting matter or dynamical
methods).

The variance of peculiar velocity ${\bf v}$
and bulk flow velocity ${\bf v}_R$ (variance of peculiar velocity 
in a sphere of radius $R$) also depend on $P(k)$,
\ba
\label{eq:<v^2>}
\langle {\bf v}^2 \rangle &=& \frac{\Omega_0^{1.20}H_0^2}{2\pi^2}
\int_0^{\infty}dk\, P(k), \nonumber \\
\langle {\bf v}^2 \rangle_R &=& \frac{\Omega_0^{1.20}H_0^2}{2\pi^2}
\int_0^{\infty}dk\, P(k)\, \left[\frac{3j_1(kR)}{kR}\right]^2.
\ea

Note that $\langle \delta_H^2 \rangle_R$, $\langle (\delta\rho/\rho)^2
\rangle_R$, $\langle \delta_{\Omega}^2 \rangle_R$, and 
$\langle {\bf v}^2 \rangle_R$ are all proportional to integrals 
over $P(k)$ multiplied by a window function $W(kR)$;
Fig.1 shows these window functions. 
Note that the window function of $\langle \delta_H^2\rangle_R$ is
somewhat similar to that of $\langle (\delta\rho/\rho)^2
\rangle_R$ and $\langle \delta_{\Omega}^2 \rangle_R$, but
very different from that of $\langle {\bf v}_R^2 \rangle$.
Also, $\langle \delta_H^2 \rangle_R$, $\langle (\delta\rho/\rho)^2
\rangle_R$, and $\langle \delta_{\Omega}^2 \rangle_R$ all decrease 
much faster with $R$ (the radius of the observed volume) than 
$\langle {\bf v}_R^2 \rangle$
(note that we have factored $1/R^2$ out of the integrals
for $\langle \delta_H^2 \rangle_R$, $\langle (\delta\rho/\rho)^2
\rangle_R$, and $\langle \delta_{\Omega}^2 \rangle_R$), 
while  $\langle {\bf v}^2 \rangle$
is independent of the size of the observed volume.

Even though $\delta_H$ is caused by the presence of
non-zero peculiar velocities [see Eq.(1)], it has only weak
correlations with the measures of the peculiar velocity field.

\section{Theoretical Power Spectra with CMB Normalization}

Let us consider cosmological structure formation models which simultaneously
satisfy constraints from the observed LCRS power spectrum (\cite{Lin96}), 
the Hubble's constant range
of $0.5 \la h \la 0.8$, cluster abundance results, and the
reasonable assumption that LCRS galaxies are approximately unbiased
on large scales relative to the mass normalization provided by the 4 year COBE DMR data. 
Following \cite{Lin96}, we assume that in the linear regime
\be
P_{galaxy,redshift\, space}(k) \simeq
b^2\, \left(1+ \frac{2}{3}\, \frac{\Omega_0^{4/7}}{b}+
 \frac{1}{5}\, \frac{\Omega_0^{8/7}}{b^2} \right)\, 
P_{mass,real \, space,linear}(k).
\ee
Three viable models are: (1) TCDM: 
$\Omega_0=1$, $n=0.7$, $h=0.5$, $\Omega_b=0.05$, $b=1.3$;
(2) $\Lambda$CDM: $\Omega_0=0.5=\Omega_{\Lambda}$, $n=1$, $h=0.5$, 
$\Omega_bh^2=0.015$, $b=0.9$;
(3) OCDM: $\Omega_0=0.5$, $n=1$, $h=0.65$, $\Omega_bh^2=0.015$, $b=0.9$.

The power spectrum of a model is given by
\be
\frac{P(k)}{h^{-3}\mbox{Mpc}^3} = P_0 \, \left(\frac{k}{\mbox{Mpc}^{-1}h}
\right)^n\, T^2(q),
\ee
where $P_0 \equiv 2\pi^2 \left(10^5\delta_{hor} \cdot 90 \right)^2
\cdot 3000^{n-1}$. The 4 year COBE DMR data give (\cite{Bunn97})
\be
10^5\delta_{hor}=\left\{ \begin{array}{ll}
1.94\, \Omega_0^{-0.785-0.05\,\ln\Omega_0}\, \exp[-0.95\,(n-1)-0.169\,(n-1)^2],
\hskip 0.5cm \Omega=1; \\
1.95\, \Omega_0^{-0.35-0.19\,\ln\Omega_0}, \hskip 0.5cm
\Omega<1, n=1.
\end{array}
\right.
\ee
$T(q)$ is the CDM transfer function, given by (\cite{Bardeen86})
\be
T(q) = \frac{\ln(1+2.34\, q)}{2.34\,q}\, \left[1+ 3.89\,q+(16.1\,q)^2
+(5.46\, q)^3+ (6.71\, q)^4 \right]^{-1/4},
\ee
where $q=k/(h\Gamma)$, $\Gamma$ is the shape parameter. We use
(\cite{HuSugi96})
\be
\Gamma = \Omega_0 h\, \alpha^{1/2} \Theta^{-2}_{2.7},
\ee
where $\Theta_{2.7}= 2.726K/2.7K$, and
\ba
\alpha &= &a_1^{-\Omega_b/\Omega_0} a_2^{-(\Omega_b/\Omega_0)^3},
\nonumber \\
a_1 &=& (46.9\, \Omega_0h^2)^{0.670}\left[
1+(32.1\, \Omega_0h^2)^{-0.532} \right], \nonumber \\
a_2 &=& (12.0\, \Omega_0h^2)^{0.424}\left[
1+(45.0\, \Omega_0h^2)^{-0.582} \right].
\ea
For the three models considered, $\Gamma=0.467$ (TCDM), 0.223 
($\Lambda$CDM), and 0.299 (OCDM) respectively.

Figs.2-5 show $\langle \delta_H^2 \rangle_R$, 
$\langle (\delta\rho/\rho)^2 \rangle_R$, $\langle \delta_{\Omega}^2 \rangle_R$, 
and $\langle {\bf v}^2 \rangle_R$, as functions of $R$, for the three
models (TCDM, $\Lambda$CDM, and OCDM).  Figure 2 indicates, for example,
that a {\it perfect} measurement of the local expansion rate in a sphere
of diameter 20,000 km/s (200$\,h^{-1}$Mpc)
would provide a 95 percent confidence interval
determination of the global value of $H_0$ with a width of 3 to 6 percent,
depending on the adopted structure formation model.

\section{Smooth Power Spectra with a Feature on Unobserved Scales}

The COBE observations of CMB fluctuations on scales larger
than 7$^o$(\cite{Bennett96}) directly probe the potential
fluctuations on scales larger than 300 $h^{-1}$ Mpc.  It is possible
to smoothly connect the COBE measurements to observations of large scale structure.  In fact, there is a family of cold dark matter models
(tilted CDM, open CDM, Lambda-dominated CDM, mixed dark matter) that
are consistent with both measurements of large-scale structure
on scales of $\sim 1 - 60$ Mpc and CMB observations on large
scales (\cite{Ratra97,Bunn97}).

There are, however, a number of observations that suggest
that the amplitude of fluctuations on the $60 - 300$ $h^{-1}$ Mpc
scale exceeds that predicted by CDM models or by smooth interpolations
between the scales probed by galaxy surveys and the COBE observations:
Lauer \& Postman (1994) measure much larger bulk flows than
predicted by standard cosmological models (\cite{Strauss95b});
deep pencil beam surveys detect excess power on scale of $\sim 100/h$
Mpc (\cite{Broadhurst90}, \cite{Cohen96}); 
two-dimensional measures of the power
spectrum in the Las Campanes redshift survey also find excess
power on the 100/$h$ Mpc scale (\cite{Landy96});
and K-band galaxy counts can be interpreted as indicating a very 
large scale local underdensity region (\cite{Phillips97}).
If the ratio of the baryon density to the closure density, 
$\Omega_b h^2$, is close to the low value suggested
by the  D/H measurements of Songaila, Wampler \& Cowie (1997), then recent
CMB measurements on the 1-2 degree scale (\cite{Netterfield97}) would
also imply excess power near the 100/$h$ Mpc scale.  
Shi et al. (1996) have used the degree scale CMB measurement
to constrain a possible feature in $P(k)$ beyond about 100$\,$Mpc
scale; however, even if
$\Omega_b h^2$ is large, it is difficult to use CMB observations
to place an upper limit on density fluctuations on this scale as
CMB fluctuations on scales smaller than that probed by
COBE may have been suppressed by reionization of the intergalactic
medium at $z > 20$.

These observations suggest that $P(k)$ is not smooth, but may
have some feature on the $60 - 300$ $h^{-1}$ Mpc scale.  
There are several interesting physical processes acting on these scales
because of the horizon size at the matter-radiation transition 
and at baryon-photon decoupling,
so it is quite possible that there is new physics acting
on these scales.  Since we are trying to constrain the uncertainties
in the Hubble's constant due to large scale structure, we
will study the possibility of excess power by adding
 a delta-function bump to a smooth matter power spectrum 
$P_0(k)$ given by a viable cosmological model. Let us write
\be
\label{eq:P(k)}
P(k)=P_0(k) + A \delta(k-k_0), 
\ee
We take $P_0(k)$ given by the tilted CDM model (TCDM)
as a convenient smooth functional form,
with $n=0.7$, $\Omega_0$=1, $\Omega_b=0.05$, and $h=0.5$.
Since $\int_0^{\infty}{\rm d}k\, P_0(k) = 828.808\, h^{-2}$Mpc$^2$,
and $\Atil =\Omega_0^{1.20} H_0^2/(2\pi^2)\left[\int_0^{\infty} 
{\rm d}k\, P_0(k)+A\right]$ is less than $(1831\,\mbox{km/s})^2$
at 95\% CL (see the following section for more details), we find that 
$A/(2\pi^2)\leq 293.3 \, h^{-2}$Mpc$^2$ at 95\% CL. 
Calculations show that the delta function spike has the most 
significant effect on the bulk flow velocity $\sqrt{\langle 
{\bf v}^2 \rangle_R}$.

Now we compute the probability distribution of $\delta_H$ measured
in a sphere of radius $R$, given the CMB dipole velocity of $v$=627 km/s. 
Using the Bayesian theorem, we find
\be
P(\delta_H|v)_R= \sum_{\Atil} P(\delta_H|\Atil)_R P(\Atil|v)
\propto \sum_{\Atil} P(\delta_H|\Atil)_R P(v|\Atil) P(\Atil),
\ee
where
\be
\label{eq:P(delH)}
P(\delta_H|\Atil)_R = \frac{1}{\sqrt{2\pi \langle \delta_H^2\rangle_R} }
\, \exp\left(- \frac{\delta_H^2(R)}{ 2\langle \delta_H^2\rangle_R} \right).
\ee
$\langle \delta_H^2\rangle_R$ is a complicated function of $\Atil$.
For the power spectrum of Eq.(\ref{eq:P(k)}), we have
\ba
\langle {\bf v}^2 \rangle & = & \Atil = \alpha_1+ \alpha_2 A,\nonumber\\
\langle \delta_H^2 \rangle_R & = & \beta_1(R)+ \beta_2(R,k_0) A.
\ea
We have defined
\ba
&&\alpha_1 \equiv \frac{\Omega_0^{1.20} H_0^2}{2\pi^2} \int_0^{\infty}
dk\, P_0(k), \hskip 1cm
\alpha_2 \equiv \frac{\Omega_0^{1.20} H_0^2}{2\pi^2},\nonumber\\
&&\beta_1 \equiv \frac{\Omega_0^{1.20}}{2\pi^2 R^2} \int_0^{\infty}
dk\, P_0(k)\, [(kR) {\cal L}(kR)]^2, \nonumber \\
&&\beta_2 \equiv \frac{\Omega_0^{1.20}}{2\pi^2 R^2} [(k_0 R) {\cal L}(k_0 R)]^2.
\ea

The parameter $A$ characterizes the departure of the power spectrum 
from the underlying smooth power spectrum; we expect the probability 
of $A$ to decrease with increasing $A$. Let us use the prior probability
\be
P(A) = \left\{ \begin{array}{ll}
1/A, \hskip 1cm A \geq A_c; \\
0, \hskip 1cm A < A_c. \end{array} \right.
\ee
\newcounter{bean}
$A_c$ is a cut-off motivated by physical considerations.
We consider two choices for the cut-off $A_c$:
\begin{list}%
{(\arabic{bean})}{\usecounter{bean}}
\item{$A_c= \alpha_1/\alpha_2 =\int_0^{\infty} {\rm d}k\, P_0(k)$;}
\item{$A_c= 0.1\,\alpha_1/\alpha_2 =0.1\,\int_0^{\infty} {\rm d}k\, 
P_0(k)$.}
\end{list}
The first choice of $A_c$ indicates the reasonable upper limit of $A_c$,
while the second choice indicates a typical value of $A_c$ of interest.

The probability distribution of $\delta_H$ given the value of $v$ is
\be
\label{eq:P(delH|v)}
P(\delta_H|v)_R=\frac{1}{\Ncal \sqrt{2\pi\beta_1}}
\int_0^{x_c} {\rm d}x\, \frac{x}{(x+1)^{3/2}
(x+ x_1)^{1/2} } \, \exp\left\{ -\left[
\frac{3v^2 \, x}{2\alpha_1 (x+1)} +
\frac{\delta_H^2 \, x}{2\beta_1 (x+ x_1)} 
\right] \right\}.
\ee
where 
\be
x_c\equiv \frac{\alpha_1}{\alpha_2}\, \frac{1}{A_c}, \hskip 1cm
x_1 \equiv \frac{\alpha_1\beta_2}{\alpha_2\beta_1},
\ee
and
\be
\Ncal \equiv \int_0^{x_c} {\rm d}x\, \frac{x^{1/2}}
{(x+1)^{3/2}} \, \exp\left\{
 -\frac{3v^2 \, x}{2\alpha_1 (x+1)} \right\}
\ee

The variance of $\delta_H$ given $v=627$ km/s is
\be
\label{eq:delH^2|v}
\langle \delta_H^2|v\rangle_R = 
\frac{\beta_1}{\Ncal}
\int_0^{x_c} {\rm d}x\, \frac{x+ x_1}
{x^{1/2} (x+1)^{3/2} } \, \exp\left\{ -
\frac{3v^2 \, x}{2\alpha_1 (x+1)} \right\}.
\ee

Figs.6-8 show $P(\delta_H|v)_R$ for $k_0=0.01\,h\,$Mpc$^{-1}$ and
$R=40, 100, 500\,h^{-1}$Mpc respectively. The solid and dashed lines are
the distributions given by Eq.(\ref{eq:P(delH|v)}), 
with $x_c=10$ ($A_c=0.1\,\alpha_1/\alpha_2$) and
$x_c=1$ ($A_c=\alpha_1/\alpha_2$) respectively;
the dotted lines are Gaussian distributions with the same variance [given
by Eq.(\ref{eq:delH^2|v})] for $x_c=10$.
Note that $P(\delta_H|v)_R$ becomes increasingly non-Gaussian 
for increasing $R$.
Table 1 lists the 95\% CL upper limits on $|\delta_H|$, $\delta_H^0$,
as well as $2\sigma\equiv 2 \sqrt{\langle \delta_H^2|v\rangle_R }$, the 95\% CL upper limit
on $\delta_H$ if its distribution were Gaussian.

\begin{table}[h]
\begin{center}
\begin{tabular}{|c|c||p{1in}|p{1in}|p{1in}|}
\hline
 & $R$ & $40\mbox{Mpc}\,h^{-1}$ & $100\mbox{Mpc}\,h^{-1}$ & 
$500\mbox{Mpc}\,h^{-1}$\\ \hline\hline
$x_c=10$ &
$\begin{array}{ll}\delta_H^0\\ 2\sigma\end{array}$
& $\begin{array}{ll} 0.115 \\ 0.120 \end{array}$ 
&  $\begin{array}{ll} 5.50\times 10^{-2} \\6.13\times 10^{-2} \end{array}$
 &  $\begin{array}{ll}9.44\times 10^{-3} \\1.05\times 10^{-2}\end{array}$ 
\\ \hline
$x_c=1$ & $\begin{array}{ll}\delta_H^0 \\2\sigma\end{array}$ 
&  $\begin{array}{ll}0.135 \\0.143 \end{array}$ 
&  $\begin{array}{ll}8.67 \times 10^{-2} \\9.56 \times 10^{-2}\end{array}$
 &  $\begin{array}{ll}1.59 \times 10^{-2}\\1.75\times 10^{-2}\end{array}$ 
\\ \hline
\end{tabular}
\end{center}
\end{table}

It is not surprising that our results depend on the prior
probability $P(A)$ via the cut-off $A_c$, since different choices
of $A_c$ represent different input of physical information.
It is worth noting that the 95\% CL upper limit on $|\delta_H|$
is close to 10\% for $R \geq 100\,$Mpc\,$h^{-1}$ with the 
largest reasonable $A_c$ (when the contribution to the mean square
peculiar velocity by the bump in $P(k)$ 
is greater or equal to the contribution by the underlying smooth
power spectrum).

Also note that in our Bayesian statistics, the measured CMB dipole velocity
of $v=627\,$km/s has the effect of distorting the distribution of
$\delta_H$ away from Gaussian by increasing the probability of smaller
$\delta_H$.

\section{A Robust Upper Limit on the Variance of $\delta_H$}

The CMB dipole velocity of $v\equiv |{\bf v}|$=627 km/s is the Galaxy's
peculiar velocity with respect to the CMB rest frame.
Note that ${\bf v}$ has three components which are three
independent Gaussian random variables.
In the previous section, it provided us with a constraint on possible
unobserved features in $P(k)$, but we can in fact use it in a more fundamental
way to provide a single integral constraint on $P(k)$ which is related
in a simple way to possible expansion rate variations, as described below.
We can infer
the distribution of $\mbox{\mbox{\~{A}}} \equiv \langle {\bf v}^2 \rangle$
by using the Bayesian theorem, 
\be
P(\mbox{\~{A}}|v) = \frac{ P(v|\mbox{\~{A}}) P(\mbox{\~{A}})}{ P(v)} \propto 
P(v|\mbox{\~{A}}) P(\mbox{\~{A}}).
\ee
$P(\mbox{\~{A}})$ is the prior probability of $\mbox{\~{A}}$. Since $\mbox{\~{A}}$/3 
is the variance of the Gaussian variable $v_x$ ($x$ component
of ${\bf v}$), it is most reasonable to choose
$P(\mbox{\~{A}}) = 1/\mbox{\~{A}}$. For diagnostic on
the dependence on the prior, let us write
\be
P(\mbox{\~{A}}) = \left\{ \begin{array}{ll}
1/\mbox{\~{A}}, \hskip 1cm \Atil \geq \Atil_c; \\
0, \hskip 1cm \Atil < \Atil_c. \end{array} \right.
\ee
$\Atil^{1/2} < \Atil_0^{1/2}$ at 95\% CL, with $\Atil_0^{1/2}$ given
by
\be
\int_0^{\Atil_0} P(\Atil|v)\, {\rm d} \Atil =0.95.
\ee
Using
\be
P(v|\mbox{\~{A}}) \propto \frac{v^2}{\mbox{\~{A}}^{3/2}} \, 
\exp\left(-\frac{3v^2}{2 \Atil} \right),
\ee
we find that $\Atil_0^{1/2}=1831\,$km/s for $\Atil_c^{1/2} \leq 200\,$km/s,
and $\Atil_0^{1/2}=2183\,$km/s for $\Atil_c^{1/2} = v=627\,$km/s.
Since $\Atil_0^{1/2}$ changes by less
than 20\% for the significant cut-off of $\Atil_c^{1/2}=v$,
we can take $\Atil_c^{1/2}=0$.

Using Eq.(\ref{eq:<v^2>}) and $\Atil_0^{1/2}=1831\,$km/s, we find 
\be
\frac{\Omega_0^{1.20}}{2\pi^2}\int_0^{\infty} \frac{dk}{h \, \mbox{Mpc}^{-1}} \,
\frac{P(k)}{h^{-3} \mbox{Mpc}^3} \leq  \left(\frac{1831\,\mbox{km/s}}
{100 \,\mbox{km/s}} \right)^2 = 335.3
\ee
at 95\% CL.

Using Eqs.(\ref{eq:<dH^2>}), (\ref{eq:<drho^2>}), and (\ref{eq:<dOmega^2>}),
we obtain at 95\% CL
\ba
\langle \delta_H^2(R) \rangle & < & \left( \frac{10.5 \,h^{-1}
\mbox{Mpc} }{R}\right)^2,
\nonumber \\
\left\langle \left(\frac{\delta\rho}{\rho}\right)^2_R \right\rangle 
&< & \Omega_0^{-1.20} \left( \frac{24.0 \,h^{-1}\mbox{Mpc}}{R}\right)^2,
\nonumber \\
\left\langle \delta_{\Omega}^2(R) \right\rangle 
&< & \Omega_0^{-1.20} \left( \frac{43.9 \,h^{-1}\mbox{Mpc}}{R}\right)^2.
\ea
We have used $[x {\cal L}(x)]^2 \leq 0.3282$, $x^2 (3j_1(x)/x)^2 \leq 1.7123$,
and $x^2[3j_1(x)/x - 2{\cal L}(x)]^2 \leq 5.7579$.

This calculation provides us with the most robust results, both in the sense 
of being model independent and of relying only on the largest and most 
observationally secure CMB anisotropy (the dipole).  They are thus also the
weakest quantitatively.  For example, 10\% variations in the expansion rate
are allowed on scales of 20,000 km/s (200$\,h^{-1}$Mpc) in diameter 
at 95 percent confidence.

\section{Summary}

Cosmologists attempt to derive properties of the large-scale structure
of space-time from local observations. These extrapolations rely
on the assumption that we are probing a fair sample of the universe,
so that physical quantities measured locally such as the Hubble's constant,
or the mass-to-light ratio are representative of the universe
as a whole.

Large-scale structure generates deviations from the Hubble flow; thus,
it is important that measurements of the Hubble's constant probe
a large enough volume so that the effects of peculiar motions are
small (\cite{Turner92}).  In this paper, we have estimated the
expected variance in the Hubble's constant due to large scale
structure.  We began by considering currently fashionable models, which
are consistent with galaxy surveys on small scales and CMB observations
on large scales.  In these models, measurements of the Hubble's
constant that are based on galaxies  
in a sphere of diameter 200$\,h^{-1}$Mpc are likely to be
within $3-6$\% (95\% confidence interval) of the global value,
and of order $1-2$\% in a sphere of 400$\,h^{-1}$Mpc diameter.
These limits assume that  the current set of large-scale structure
models are good approximation to the primordial power spectrum.

The CMB dipole velocity (the velocity of the Galaxy with respect
to the CMB rest frame) places a strong constraint
on the amplitude of large scale structure. 
If there were enormous local void and density fluctuations, 
as suggested by several authors (\cite{Harrison93,Wu95,Wu96}), 
then we would expect that the Galaxy would be moving 
with respect to the microwave background rest frame
at a velocity much larger than the meaured value of $627\pm22\,$km/s
(\cite{Kogut93,Fixsen94}).  
Thus, the measured CMB dipole velocity can be used to 
derive strong constraints on density fluctuations on scales 
larger than those probed by redshift surveys.
We have used these constraints to calculate the variations in the
Hubble's constant, $\delta_H$, for power spectra which have a sharp bump
on unobserved scales; we find that the 95\% CL upper limit of
$|\delta_H|$ increases approximately by a factor of two.
 
Finally, we have used the constraints derived from the CMB dipole
velocity to place a robust limit on variations in the Hubble's constant.  
With the CMB dipole measurement alone, we are able to constrain variations in
the Hubble's constant to be less than 10\% on scales of 20,000 km/s 
(200$\,h^{-1}$Mpc) in diameter at 95 percent confidence.  
Thus, measurements of H$_0$ that probe out
to this scale are likely to be accurately probing the global
expansion rate of the universe.

\section{Acknowledgements}

Y.W. and E.L.T. acknowledge support from NSF grant AST94-19400.
DNS acknowledges support from the MAP/MIDEX project.


\clearpage

\figcaption[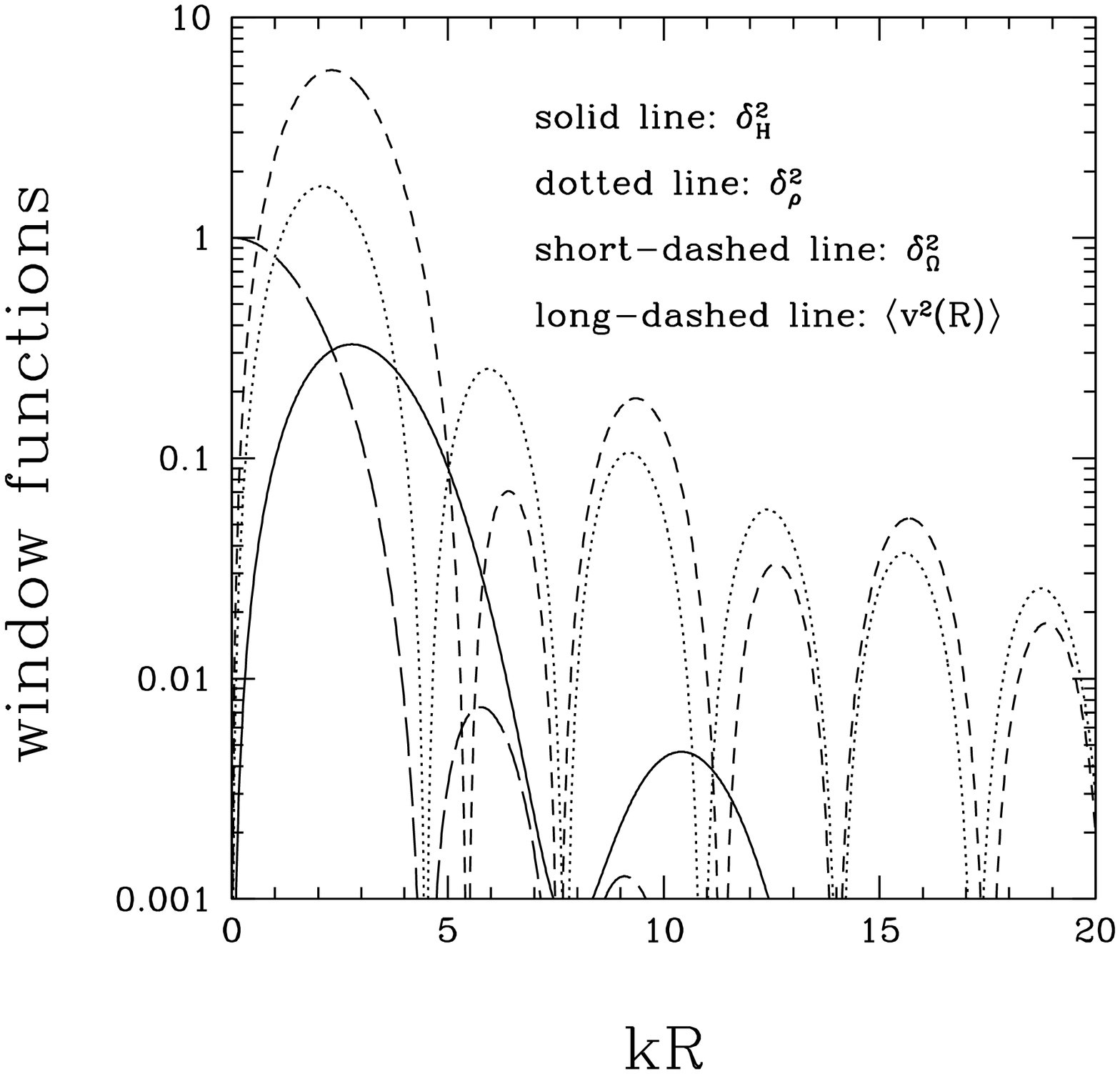]{Window functions of
$\langle \delta_H^2 \rangle_R$ (solid line), $\langle (\delta\rho/\rho)^2
\rangle_R$ (dotted line), $\langle \delta_{\Omega}^2 \rangle_R$ (short-dashed
line), and $\langle {\bf v}^2 \rangle_R$ (long-dashed line).}

\figcaption[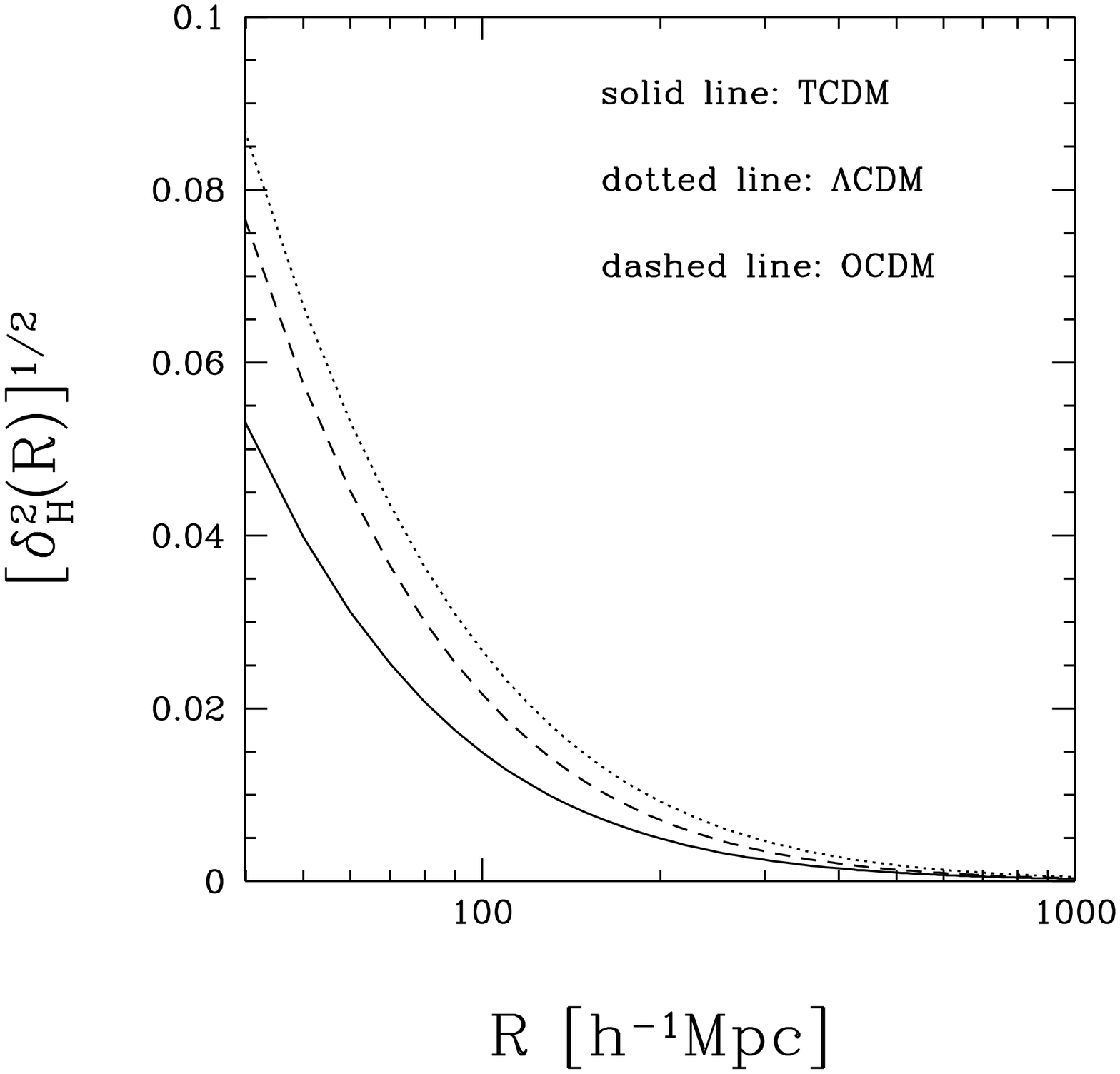]
{$\langle \delta_H^2 \rangle_R$ as function of $R$, for the three
models (TCDM, $\Lambda$CDM, and OCDM).}

\figcaption[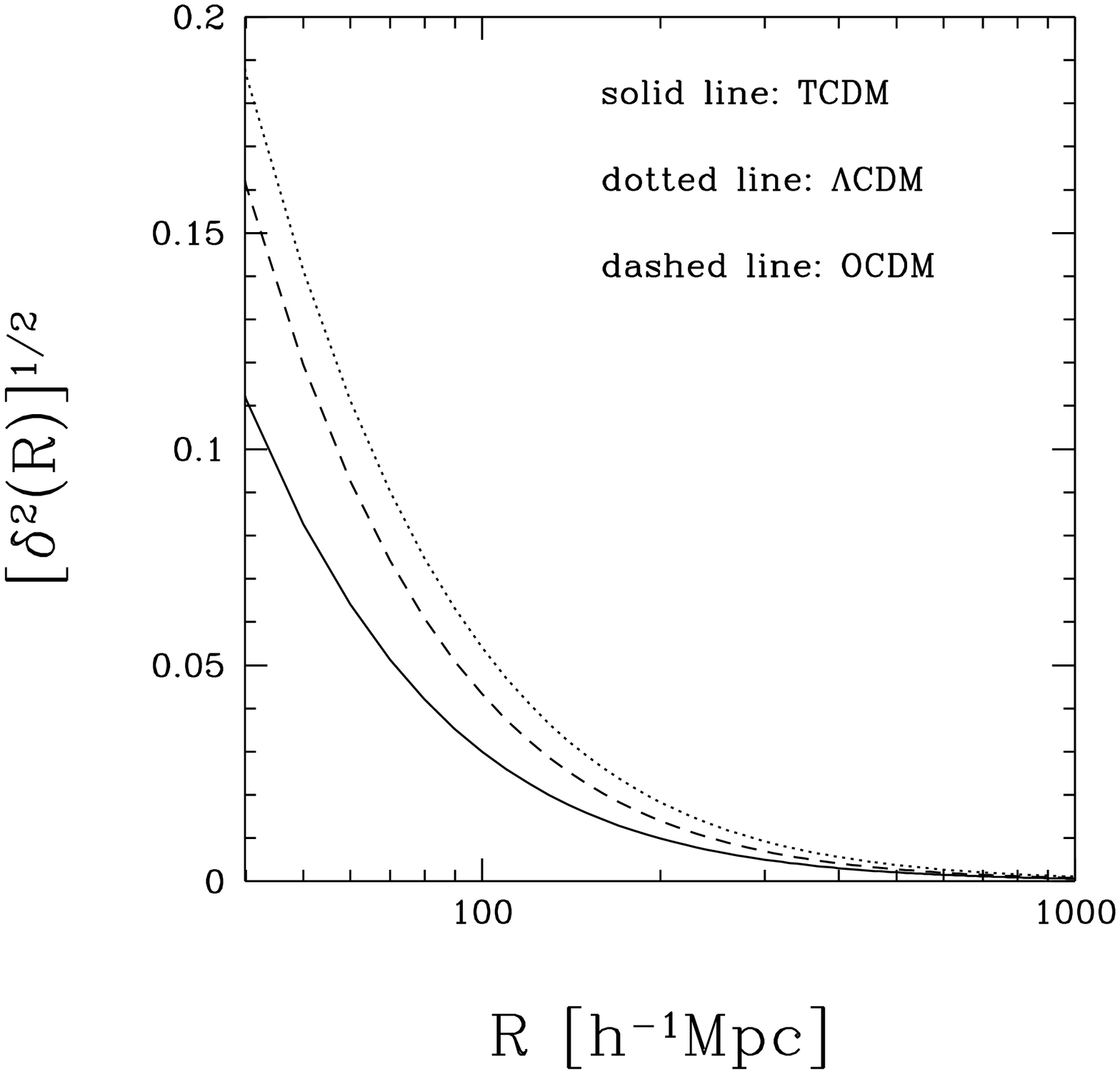]
{$\langle (\delta\rho/\rho)^2 \rangle_R$ as function of $R$, for the three
models (TCDM, $\Lambda$CDM, and OCDM).}

\figcaption[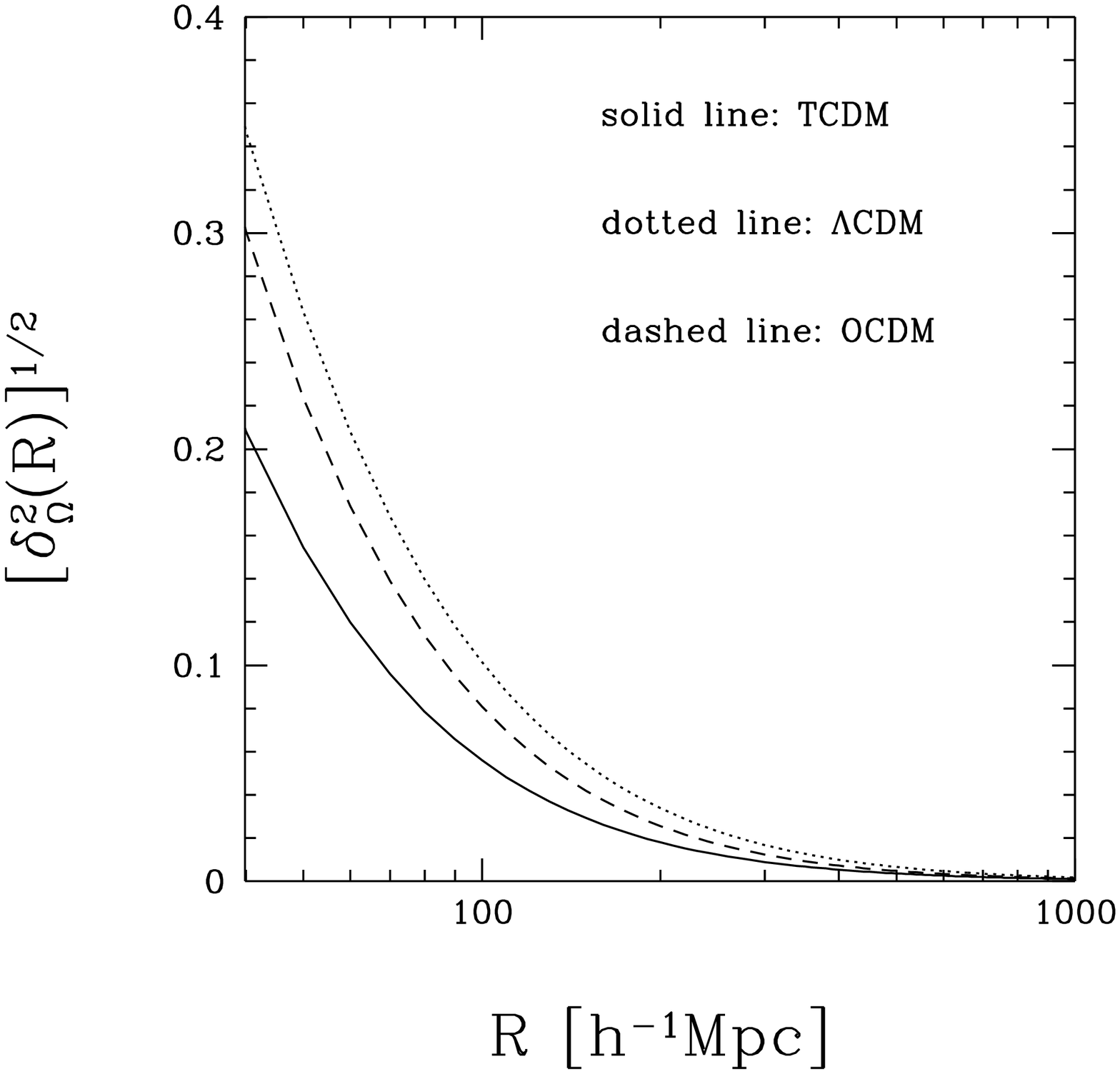]
{$\langle \delta_{\Omega}^2 \rangle_R$ as function of $R$, for the three
models (TCDM, $\Lambda$CDM, and OCDM).}

\figcaption[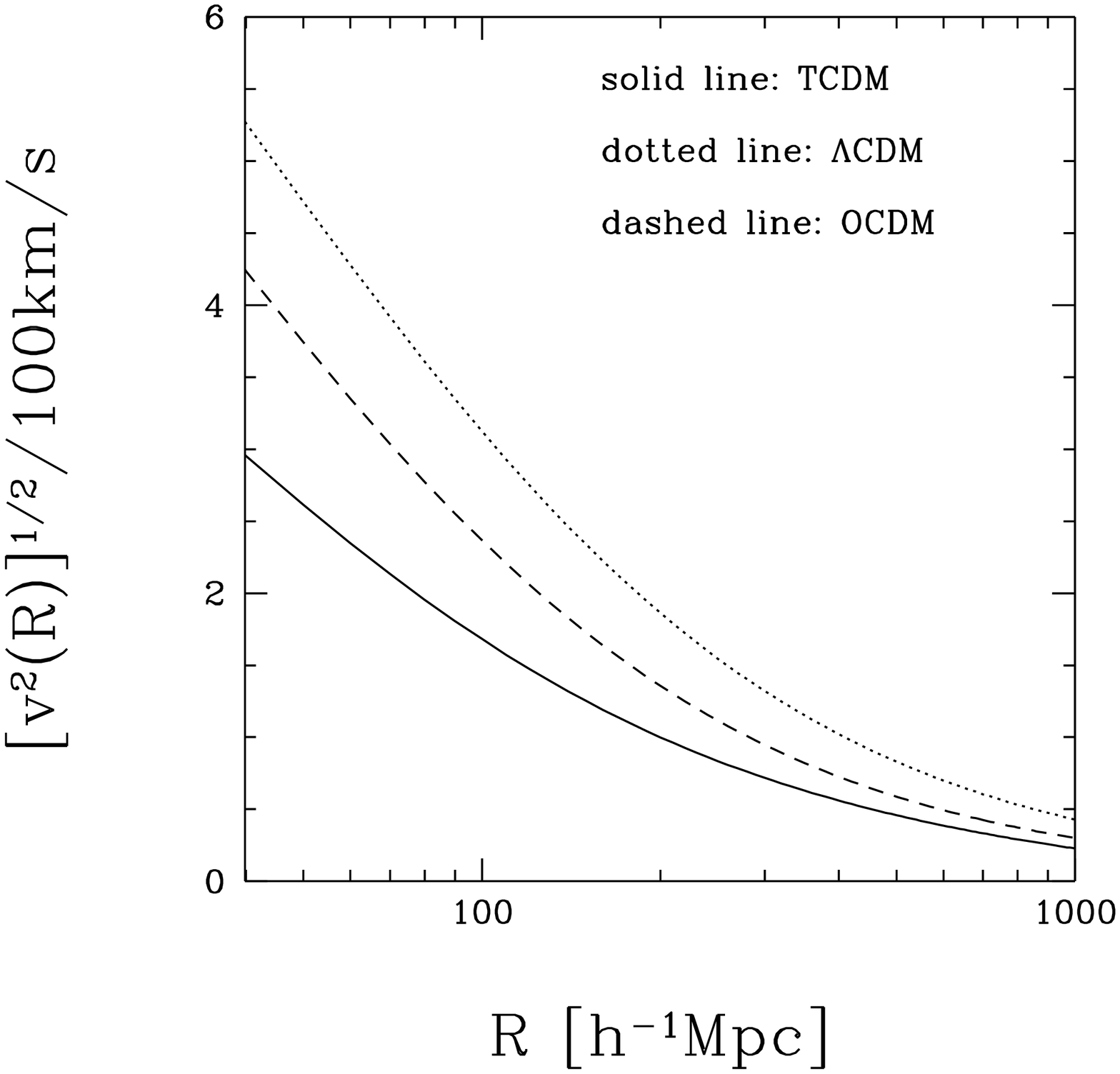]
{$\langle {\bf v}^2 \rangle_R$ as function of $R$, for the three
models (TCDM, $\Lambda$CDM, and OCDM).}

\figcaption[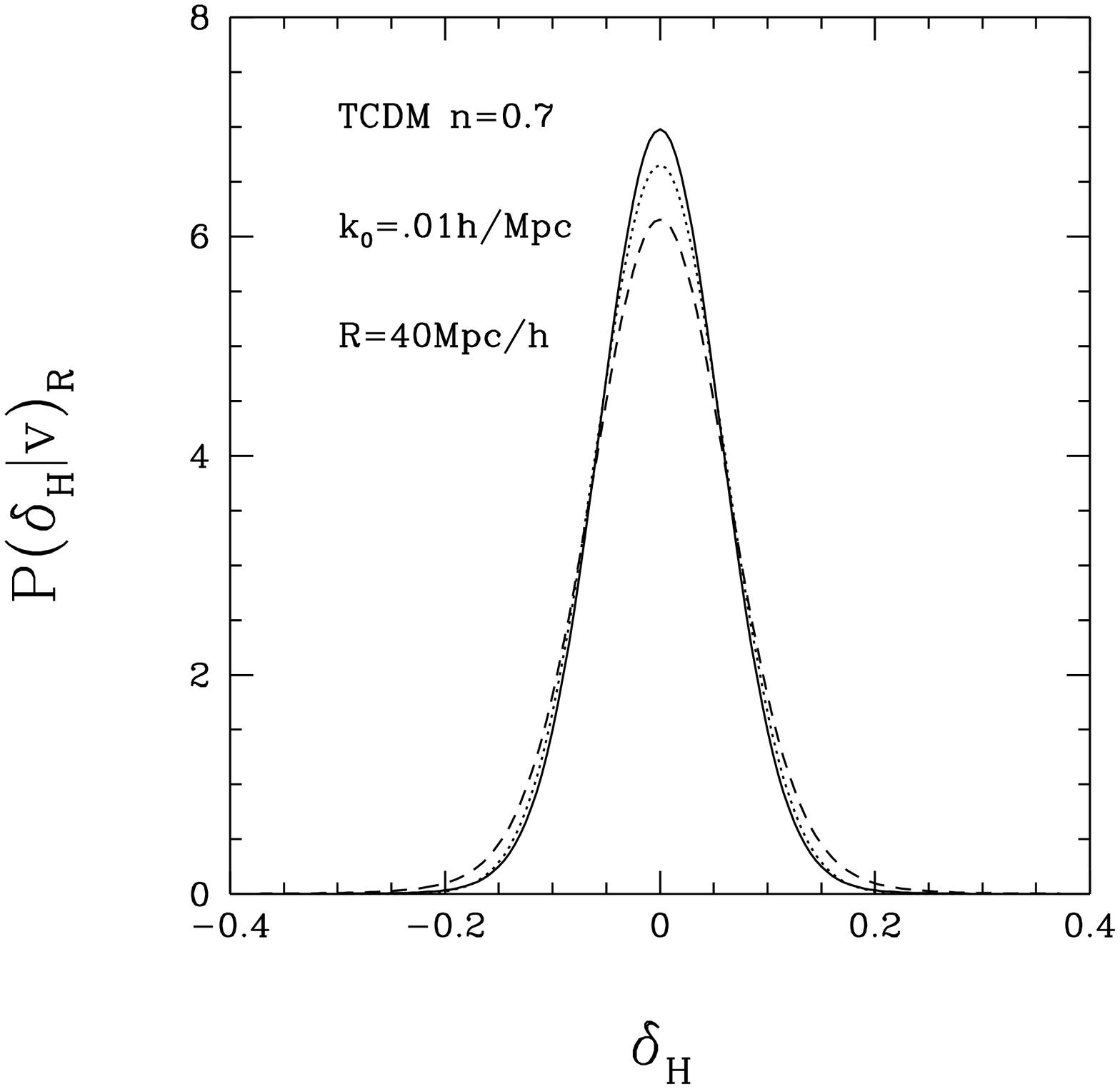]
{$P(\delta_H|v)_R$ for $k_0=0.01\,h$Mpc$^{-1}$ and
$R=40\,h^{-1}$Mpc. The solid and dashed lines are
the distributions given by Eq.(\ref{eq:P(delH|v)}), 
with $x_c=10$ ($A_c=0.1\,\alpha_1/\alpha_2$) and
$x_c=1$ ($A_c=\alpha_1/\alpha_2$) respectively;
the dotted lines are Gaussian distributions with the same variance [given
by Eq.(\ref{eq:delH^2|v})] for $x_c=10$.}

\figcaption[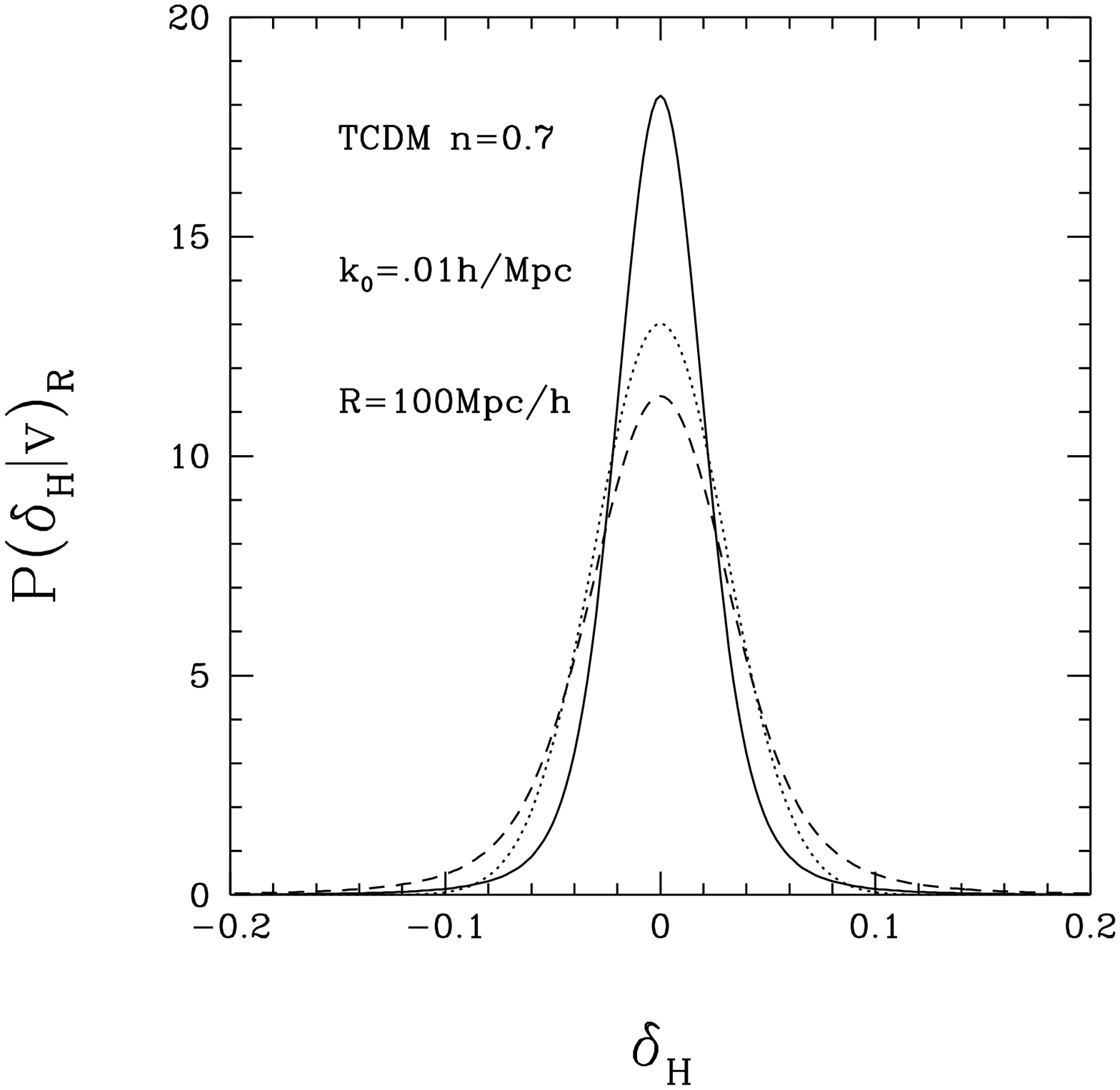]
{Same as Fig.6, for $R=100\,h^{-1}$Mpc.}

\figcaption[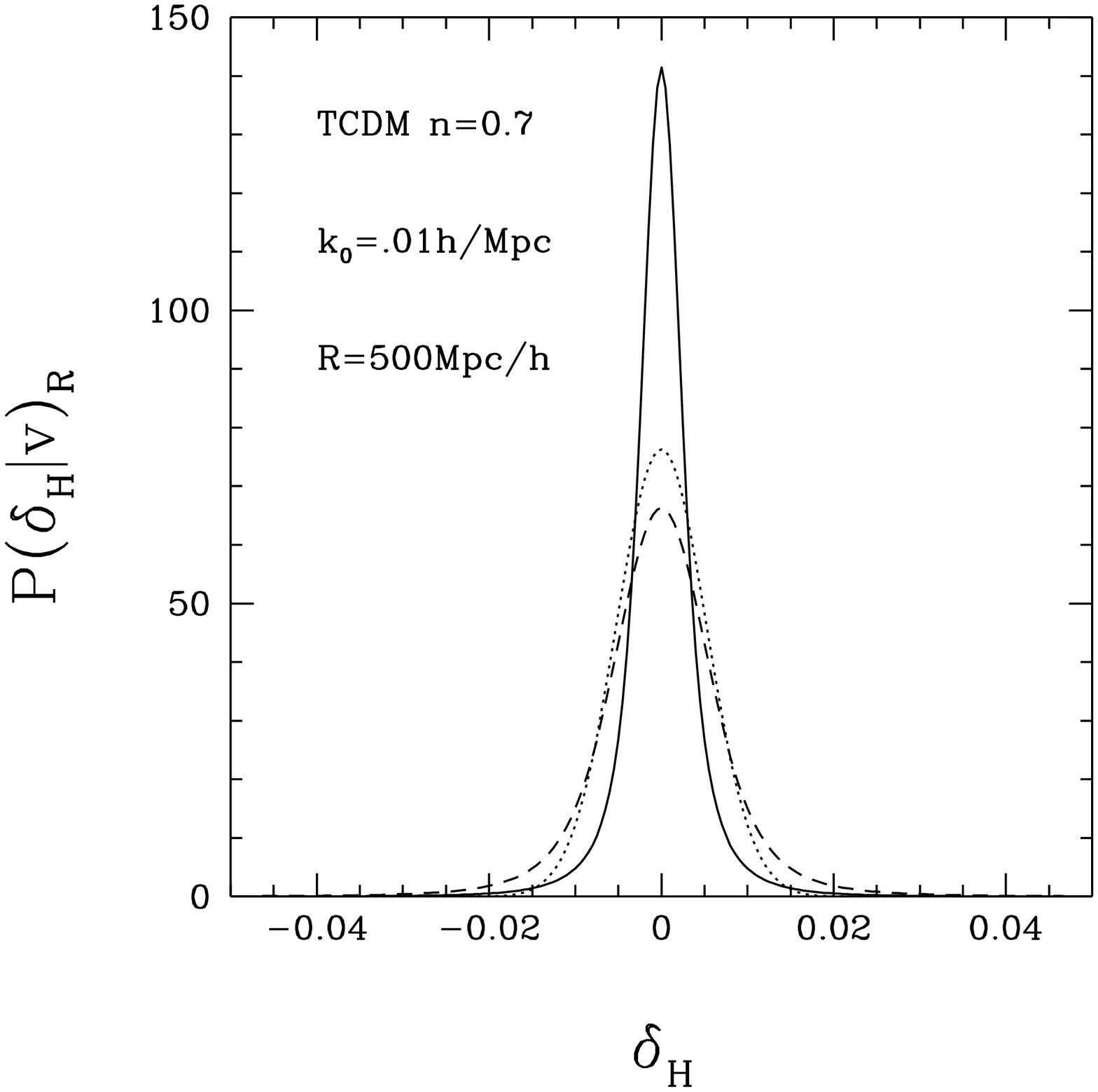]
{Same as Fig.6, for $R=500\,h^{-1}$Mpc.}

\clearpage

\setcounter{figure}{0}
\plotone{fig1.eps}
\figcaption[fig1.eps]{Window functions of
$\langle \delta_H^2 \rangle_R$ (solid line), $\langle (\delta\rho/\rho)^2
\rangle_R$ (dotted line), $\langle \delta_{\Omega}^2 \rangle_R$ (short-dashed
line), and $\langle {\bf v}^2 \rangle_R$ (long-dashed line).}

\plotone{fig2.eps}
\figcaption[fig2.eps]
{$\langle \delta_H^2 \rangle_R$ as function of $R$, for the three
models (TCDM, $\Lambda$CDM, and OCDM).}

\plotone{fig3.eps}
\figcaption[fig3.eps]
{$\langle (\delta\rho/\rho)^2 \rangle_R$ as function of $R$, for the three
models (TCDM, $\Lambda$CDM, and OCDM).}

\plotone{fig4.eps}
\figcaption[fig4.eps]
{$\langle \delta_{\Omega}^2 \rangle_R$ as function of $R$, for the three
models (TCDM, $\Lambda$CDM, and OCDM).}

\plotone{fig5.eps}
\figcaption[fig5.eps]
{$\langle {\bf v}^2 \rangle_R$ as function of $R$, for the three
models (TCDM, $\Lambda$CDM, and OCDM).}

\plotone{fig6.eps}
\figcaption[fig6.eps]
{$P(\delta_H|v)_R$ for $k_0=0.01\,h$Mpc$^{-1}$ and
$R=40\,h^{-1}$Mpc. The solid and dashed lines are
the distributions given by Eq.(\ref{eq:P(delH|v)}), 
with $x_c=10$ ($A_c=0.1\,\alpha_1/\alpha_2$) and
$x_c=1$ ($A_c=\alpha_1/\alpha_2$) respectively;
the dotted lines are Gaussian distributions with the same variance [given
by Eq.(\ref{eq:delH^2|v})] for $x_c=10$.}

\plotone{fig7.eps}
\figcaption[fig7.eps]
{Same as Fig.6, for $R=100\,h^{-1}$Mpc.}

\plotone{fig8.eps}
\figcaption[fig8.eps]
{Same as Fig.6, for $R=500\,h^{-1}$Mpc.}

\end{document}